\documentclass[aps,prd,reprint,nofootinbib,showpacs,showkeys,superscriptaddress]{revtex4-1}

\usepackage{amsmath}
\usepackage{amssymb}
\usepackage{amsfonts}
\usepackage{amsthm}
\usepackage{graphicx}

\usepackage{latexsym}

\begin{document}

\title{Experimental limits on the free parameters  of higher-derivative gravity
}
\thanks{Contribution to the proceedings of the 14th Marcel Grossmann Meeting, Rome 12-18 July 2015.}

\author{Breno L. Giacchini}%
\affiliation{
Centro Brasileiro de Pesquisas F\'{\i}sicas
\\
Rio de Janeiro, RJ, Brazil
\\
breno@cbpf.br
}%

\date{December 30, 2015}

\begin{abstract}
Fourth-derivative gravity has two free parameters, $\alpha$ and $\beta$, which couple the curvature-squared terms $R^2$ and $R_{\mu\nu}^2$. Relativistic effects and short-range laboratory experiments can be used to provide upper limits to these constants. In this work we briefly review both types of experimental results in the context of higher-derivative gravity. The strictest limit follows from the second kind of test. Interestingly enough, the bound on $\beta$ due to semiclassical light deflection at the solar limb is only one order of magnitude larger.
\end{abstract}


\maketitle


\section{Introduction}

Fourth-derivative gravity (HDG), \textit{i.e.} the system described by the action
\begin{equation}
S= \int d^4x \sqrt{-g}\left( \frac{2R}{\kappa^2} + \frac{\alpha}{2}R^2 - \frac{\beta}{2}R_{\mu \nu}^2  - \mathcal{L}_M \right) ,
\end{equation}
has two positive dimensionless free parameters $\alpha$ and $\beta$, which must satisfy the constraint $(3\alpha - \beta) > 0$ in order to avoid tachyons on the model~\cite{Stelle78}. Here $\kappa^2 = 32\pi G$ and $\mathcal{L}_M$ is the Lagrangian density for the usual matter. Contrary to general relativity (GR), this model is renormalizable along with its matter couplings~\cite{Stelle77}, but contains a massive spin-2 ghost which violates unitarity~\cite{Johnston88}. Nonetheless, it can be regarded as an effective theory at familiar energy scales and plays a relevant role in the search for quantum gravity~\cite{AcciolyEtAl15}.

In this sense it is important to find bounds on the free parameters $\alpha$ and $\beta$. Some attempts in this vein were carried out in the 1960s and 1970s, when interest grew on general fourth-order theories, by studying classical tests of GR and short-range laboratory experiments~\cite{Anos60e70}. The strictest limit\footnote{Throughout this work all figures associated to $\alpha$ and $\beta$ must be understood in the sense of order of magnitude.} at that time, $\alpha,\beta \leq 10^{74}$, was set by Stelle~\cite{Stelle78} using Long's laboratory test of the inverse-square law~\cite{Long}. A better upper bound on $\beta$ was later determined in~\cite{AcciolyBlas} from the analysis of light deflection in the framework of tree-level HDG, \textit{i.e.} considering gravity as a classical external field and the photon as a quantum particle. A more recent and detailed discussion on the subject can be found in Ref.~\cite{AHGH}, where this phenomenon is studied in both classical and semiclassical approaches. The constraint $\beta \leq 10^{61}$ is set within the semiclassical context, in which photon propagation depends on its energy yet in first order at the tree-level.

Both Newtonian inverse-square law experiments and relativistic effects have recently been used to establish bounds on many interesting models, such as $f(R)$ and scalar-tensor theories, extra-dimension scenarios and Standard Model extensions~\cite{varios,Scharer,Naf10,Berry11}. Our aim in this work is to review those experimental results in the realm of HDG. Natural units ($c = \hbar = 1$) are used throughout.

\section{Gravitational Red-shift in HDG}

It follows from Einstein's equivalence principle that the difference of gravitational potential between two atoms would be perceived as a disparity in the energy of the photons they emit:
\begin{equation} \label{redshift}
\sigma \equiv \frac{\nu_a - \nu_b}{\nu_a} = V(b) - V(a),
\end{equation}
where $\nu_a$ and $\nu_b$ are the frequencies of the photons emitted in $a$ and $b$ (as measured by the same observer), and $V$ is the gravitational potential.

An experimental departure from GR's prediction to gravitational red-shift could be caused either by the violation of the equivalence principle, or of the inverse-square force law~\cite{Hughes1990}. The latter occurs in the case of HDG, since in the weak field approximation ($g_{\mu\nu} \equiv \eta_{\mu\nu} + \kappa h_{\mu\nu}$) it yields the effective potential~\cite{Stelle77}
\begin{equation} \label{HDGpotential}
V(\textbf{r}) = MG \left[  -\frac{1}{r} - \frac{1}{3}\frac{e^{-m_0 r}}{r} + \frac{4}{3} \frac{e^{-m_2 r}}{r} \right]
\end{equation}
for a point-like mass $M$ resting in the origin of the coordinate system. Here we defined $m_0^2 \equiv \frac{2}{(3\alpha - \beta)\kappa^2}$ and $m_2^2 \equiv \frac{4}{\beta\kappa^2}$. Substitution of the potential \eqref{HDGpotential} into \eqref{redshift} allows the study of the gravitational red-shift in the framework of HDG.

\subsection{Stellar spectra measurements}

Measurements of gravitational red-shift in stellar spectra are ubiquitous in astronomy as a mean of finding the mass-to-radius ratio of a star, on the assumption that GR holds. Therefore, in order to use these experimental data as a gravity test, the mass and radius of the star must be accurately determined by an independent technique. In this context, apart the Sun itself, double systems with a white dwarf can be useful.

Since our distance $D$ to the star is huge in comparison to its radius $R$, the equation~\eqref{redshift} which governs the effect simplifies to
\begin{equation} \label{stellar_redshift}
\sigma = \left[ 1 + \frac{1}{3}\frac{e^{-m_0 R}}{R} - \frac{4}{3} \frac{e^{-m_2 R}}{R} \right] \sigma_\mathrm{E} ,
\end{equation}
which is the same of~\eqref{HDGpotential}, being $\sigma_\mathrm{E} = MG/R$ the shift predicted by Einstein's gravity. Insomuch as the Yukawa terms have coefficients of opposite signs, it is worthwhile to notice that the higher-order correction vanishes for
\begin{equation}
\alpha = \frac{1}{3} \left[ \beta + 2 \left( \frac{\kappa \ln 4}{R} - \sqrt{\frac{4}{\beta}}  \right) ^{-2} \right] ,
\end{equation}
provided that
\begin{equation}
\beta < \beta_\mathrm{c} \equiv \left( \frac{2R}{\kappa \ln 4} \right) ^2.
\end{equation}

This condition on $\beta$ guarantees to the $m_0$-term the capability of compensating that of $m_2$, which has a coefficient four times larger in absolute value. Of course, exact cancellation would be a circumstantial phenomenon. If $\beta$ were larger than this critical $\beta_\mathrm{c}$, the red-shift would be less than the predicted by GR, in spite of $\alpha$.

The gravitational contribution to the spectrum emitted by the Sun could only be precisely measured after the 1950s, when solar physics models were developed so as to account for the Doppler shifts due to the dynamic character of the photosphere~\cite{Blamont1961,Snider1972,LoPresto1991}. Up to now, Einstein's prediction to this phenomenon has been verified within the uncertainty of 2\%~\cite{LoPresto1991}.

On the other hand, red-shift is more prominent in white dwarfs' spectra, because of their larger mass-to-radius ratio. Smaller radii play an extra role in HDG: since the Yukawa potentials represent short-range interactions, a smaller distance allows a stronger constraint on the free parameters. Accurate results of this type can be derived from the spectra of Sirius B and 40 Eridani B, for which masses and radii are known with enough precision~\cite{Barstow2005,Popper54,Reid96}.

Table~\ref{tabelaRedShift} shows the most precise red-shift measurements for the aforementioned stars, together with the star's radius (in solar radii $R_\odot$), and the shift predicted by GR. Error bars of the theoretical $\sigma_\mathrm{E}$ are due to uncertainties on the stellar parameters\footnote{We considered, for Sirius B~\cite{Barstow2005}, $R = (8.64 \pm 0.12)\times 10^{-3} R_\odot$ and $M = (0.978 \pm 0.005) M_\odot$; and $R = (1.36 \pm 0.02)\times 10^{-2} R_\odot$ and $M = (0.50 \pm 0.01) M_\odot$ for 40 Eridani B~ \cite{Shipman97}.}. Each of these measurements can be compared with HDG's formula~\eqref{stellar_redshift}, regarded as a function $\sigma = \sigma(\alpha,\beta)$. In general, the most relevant subset of the parameter space which fits the observational data is constrained by the conditions $\beta \leq \beta_\mathrm{c}$ and $\beta_\mathrm{c} \geq \alpha > \beta/3$, the last inequality being the no-tachyon prescription. However, as $\beta$ approaches the critic $\beta_\mathrm{c}$ it is possible to have $\alpha > \beta_\mathrm{c}$, since the $m_2$-term dominates despite $m_0$.

The limiting value $\beta_\mathrm{max} \sim \beta_\mathrm{c}$ derived from each situation is reported in the last column of Table~\ref{tabelaRedShift}. The smaller radii of the white dwarfs resulted in the bound $\beta \leq 10^{81}$ which is lower than the derived from the solar spectrum, notwithstanding the precision of the latter being better than the former's.

\begin{table*}[t]
	\centering
	\begin{footnotesize}
	\caption{Measurements of red-shift at astronomical scale and limits on $\beta$}
  \begin{tabular}{@{}cccccc}
  \hline
     Method & $R/R_\odot$ & $\sigma_\mathrm{E}$ & $\sigma_{\mathrm{exp}}/\sigma_\mathrm{E}$ & Ref. & $\beta_\mathrm{max}$ \\
\hline
Sun & 1 & $2.12 \times 10^{-6}$ & 0.99 $\pm$ 0.02  & \cite{LoPresto1991} & $10^{85}$ \\
41 Eridani B & $1.36\times 10^{-2}$ & $(7.8 \pm 0.3) \times 10^{-5}$ & 1.1 $\pm$ 0.1  & \cite{Reid96}   & $10^{81}$ \\
Sirius B & $8.64\times 10^{-3}$ & $(2.41 \pm 0.05) \times 10^{-4}$ & 1.11 $\pm$ 0.09 & \cite{Barstow2005}   & $10^{81}$ \\
Pound-Rebka-Snider & $9.16 \times 10^{-3}$ & $2.454 \times 10^{-15}$ & 0.997 $\pm$ 0.008  & \cite{PoundSnider64}   & $10^{78}$ \\
\hline
		\end{tabular}
	\label{tabelaRedShift}
	\end{footnotesize}
\end{table*}

\subsection{The Pound-Rebka-Snider experiment}

Even though red-shift was conceived as an astrophysical test of gravitation, the most accurate measurements have been obtained in controlled environments, namely by the Vessot-Levine~\cite{Vessot1980} and the Pound-Rebka-Snider~\cite{PoundRebka59,PoundSnider64} experiments. The former consisted on a maser inside a spaceship which travelled up to 10,000~km above Earth's surface. The frequency of the maser was monitored along the orbit of the rocket and led to the verification of GR's result within a precision of $7\times10^{-5}$. However, comparison between this figure and the model is not straightforward, for in HDG the departure from the Einsteinian prediction is distance-dependent and the whole orbit must be modelled. This difficulty was also appointed in Ref.~\cite{Hughes1990}.

A direct comparison, though, can be carried out in the case of Pound-Rebka-Snider experiment, whose detailed description can be found in Ref.~\cite{PoundRebka59}. In short, the authors compared emission and absorption of $\gamma$ rays by nuclei located on the top and on the bottom of a 22.5-m height tower, the potential being related to Earth's gravitational field. Under these circumstances, GR predicts the shift $\sigma_\mathrm{E} = 2.454 \times 10^{-15}$; the most precise experimental result was $\sigma_{\mathrm{exp}}/\sigma_\mathrm{E}=0.997\pm0.008$~\cite{PoundSnider64}. 

The same procedure used to analyse the stellar spectra measurements can be applied here; however, the small separation between the atoms makes it necessary to use the full expression (2). Hence, two relevant characteristic lengths are present in the problem: Earth's radius and the height of the tower. This  prohibits the formal cancellation of the Yukawa terms -- which in some sense allowed $\alpha$ to be larger than $\beta_\mathrm{max}$ in the case of the stellar spectra.

It is possible to show that agreement between HDG's predictions and the measured data can occur provided that $\alpha,\beta \leq 10^{78}$, as displayed in the last row of Table~\ref{tabelaRedShift}. Notwithstanding the similitude between white dwarfs' radii and Earth's one, here the closer figures of $R$ and $D$ resulted in a more significant contribution owed by the higher-order terms. Together with a higher precision, this yields the strictest constraint on the coefficients, among current red-shift measurements.

\section{Short-range Laboratory Experiments}

Since Long's work~\cite{Long}, many advances have occurred on the field of laboratory precision experiments on gravitation, aiming to probe small and smaller distances with increasing accuracy as time goes by~\cite{NewmanBergBoynton}. Verification of the inverse-square law has now reached the micrometer scale and allowed the setting of bounds on the constants of various models which predict Yukawa or power-law interactions~\cite{varios2,Kapner07}.

In this spirit, in the non-relativistic classical realm it is possible to use the interparticle potential (3) as the elementary potential to evaluate gravitational interaction between extended bodies. The E\"ot-Wash torsion-balance experiments~\cite{Kapner07} are the most restrictive tests of a Yukawa interaction with strength compatible with HDG, \textit{i.e.}~having coefficients on the order of $1/3$ and $4/3$. Gravitational force law was probed at separations between 9.53~mm and 55~$\mu$m in that work.

Considering the correction due to only one Yukawa potential with mass $m$, Ref.~\cite{Kapner07} provides the limit $m \geq 10^{4}$~m$^{-1}$. It is remarkable that this constraint differs from the result of~\cite{AcciolyBlas,AcciolyEtAl15} by only one order of magnitude. In fact, $\beta \leq 10^{61}$ implies $m_2 \geq 10^{3}$~m$^{-1}$. The potential of HDG, however, has two Yukawa terms with opposite signs which act as a tug of war between attracting and repelling forces. Solutions with masses $m_0$ and $m_2$ smaller than $10^{4}$~m$^{-1}$ could, in principle, yield correct results. But here forces are actually measured -- differently from the stellar spectra red-shift, which probed the potential expression directly. Hence, even the vanishing of the potential at a certain separation may cause a detectable torque; and the determination of a finer constraint would require the modelling of the whole experiment taking into account the potential~\eqref{HDGpotential}. The upper limit derived from a preliminary analysis of the E\"ot-Wash experiment is, thus, $\alpha,\beta \leq 10^{60}$.

\section{Closing Remarks}

As it is long known, classical tests at astronomical scales only offer weak constraints on the coupling constants $\alpha$ and $\beta$~\cite{Stelle78}. For instance, measurements of the red-shift in the solar spectra yield the upper bound $10^{85}$, while deflection of light rays grazing the Sun leads to $10^{83}$~\cite{AHGH}. Both tests have the same characteristic length of one solar radius, the difference between the limits they convey being due to the current experimental precision which is better in the latter thanks to the use of very long baseline interferometry of radio waves~\cite{Fomalont09}.

Gravitational red-shift at the scale of one Earth radius yields the limit $10^{78}$ (Pound-Rebka-Snider experiment), which is still large enough to violate the Newtonian force law at familiar separations of centimetres and kilometres.

We did not consider in our analysis the perihelion precession, since even in the realm of GR it is only satisfactorily predicted if the full non-linear theory is used. But given that the tightest constraint would follow from the length scale of Mercury's orbit, we cannot expect a better result than the other classical tests.

The best constraints are given by short-range tests of the inverse-square law, which is intuitive since the higher-order terms are supposed to represent corrections at small distances, close to the Planck scale, where quantum gravity effects would become relevant. The most stringent bound at the present time is $\alpha,\beta \leq 10^{60}$, fourteen orders of magnitude smaller than the corresponding figures available forty years ago~\cite{Stelle78}.

It is astonishing that a constraint only one order of magnitude larger than this can be set to the parameter $\beta$ by the semiclassical analysis of the deflection of photons passing by the Sun~\cite{AHGH}. This only happens because of the $R_{\mu\nu}^2$-sector, which is responsible for both the repelling force and the energy-dependent scattering. (Light bending does not depend on $\alpha$, neither semiclassically~\cite{AcciolyBlas}.)

In general, the classical limits presented in this work agree with those found to $\alpha$ in the context of quadratic gravity (as in Refs.~\cite{Naf10,Berry11}), \textit{i.e.} only considering the sectors $R$ and $R^2$, which can be regarded as a first order expansion of $f(R)$ models. This is consequence of the similitude between the corrections supplied by the sectors $R^2$ and $R_{\mu\nu}^2$ in the classical non-relativistic domain.

\section*{Acknowledgements}

The author is very grateful to A. Accioly for the fruitful discussions; and to CAPES and ICRANet for supporting his attendance at the 14th Marcel Grossmann Meeting. The author also acknowledges CNPq for having supported his M.Sc. project (which was partially presented in this work), and now for supporting his Ph.D. studies.

\end{document}